\documentclass[3p,number,sort&compress,twocolumn]{elsarticle} %preprint, review, authoryear
\usepackage{titlesec}
\usepackage[colorlinks=true]{hyperref}
\usepackage{parskip}
\usepackage[utf8]{inputenc}
\usepackage{verbatim}
\usepackage[font=small,labelfont=bf]{caption}
\usepackage{units}
\usepackage{amsmath}
\usepackage{lineno}

\usepackage{algorithm}
\usepackage{algpseudocode}

\usepackage{subcaption}

\usepackage{tikz}
\usetikzlibrary{arrows.meta}
\usetikzlibrary{calc}

\def\centerarc[#1](#2)(#3:#4:#5)% Syntax: [draw options] (center) (initial angle:final angle:radius)
    { \draw[#1] ($(#2)+({#5*cos(#3)},{#5*sin(#3)})$) arc (#3:#4:#5); }

\makeatletter
\newenvironment{breakablealgorithm}
  {% \begin{breakablealgorithm}
   \vspace{0.5em}
   \begin{center}
     \refstepcounter{algorithm}% New algorithm
     \hrule height.8pt depth0pt \kern2pt% \@fs@pre for \@fs@ruled
     \renewcommand{\caption}[2][\relax]{% Make a new \caption
       {\raggedright\vspace{-0.6em}\textbf{\ALG@name~\thealgorithm} ##2\par}%
       \ifx\relax##1\relax % #1 is \relax
         \addcontentsline{loa}{algorithm}{\protect\numberline{\thealgorithm}##2}%
       \else % #1 is not \relax
         \addcontentsline{loa}{algorithm}{\protect\numberline{\thealgorithm}##1}%
       \fi
       \kern2pt\hrule\kern2pt
     }
  }{% \end{breakablealgorithm}
     \kern2pt\hrule\relax% \@fs@post for \@fs@ruled
   \end{center}
  }
\makeatother

\usepackage{fancyhdr}
\fancypagestyle{plain}{%
  \fancyhf{}
  \fancyhead[L]{\sc Barnes}
  \fancyhead[R]{\sc Parallel Landscape Evolution}
  \cfoot{\thepage}
  
  }
\biboptions{square}

\newcommand{\todo}[1]{}

\hypersetup{
  pdfauthor   = {Richard Barnes (ORCID: 0000-0002-0204-6040)},
  pdftitle    = {Accelerating a fluvial incision and landscape evolution model with parallelism},
  pdfproducer = {LaTeX},
  pdfcreator  = {pdfLaTeX}
}

\algblockdefx{MultiForAll}{EndMultiForAll}{\textbf{for$_\parallel$ all }}{\textbf{end for$_\parallel$}}
\algblockdefx{MultiFor}{EndMultiFor}{\textbf{for$_\parallel$ }}{\textbf{end for$_\parallel$}}

\algtext*{EndWhile} % Remove "end while" text
\algtext*{EndIf}    % Remove "end if" text
\algtext*{EndFor}   % Remove "end if" text
\algtext*{EndMultiFor}
\algtext*{EndMultiForAll}

\begin{document}
\thispagestyle{empty}

%\linenumbers

\begin{frontmatter}
  \title{\sc Accelerating a Landscape Evolution Model with Parallelism}

  \author[rb]{Richard Barnes\corref{cor_rb}}
  \ead{richard.barnes@berkeley.edu}
  \address[rb]{Energy \& Resources Group, Berkeley, USA}
  \cortext[cor_rb]{Corresponding author. ORCID: 0000-0002-0204-6040}

  \begin{abstract} %~??? words
  \noindent Solving inverse problems and achieving statistical rigour in landscape evolution models requires running \textit{many} model realizations. Parallel computation is necessary to achieve this in a reasonable time. However, no previous algorithm is well-suited to leveraging modern parallelism. Here, I describe an algorithm that can utilize the parallel potential of GPUs, many-core processors, and SIMD instructions, in addition to working well in serial. The new algorithm runs 43\,x faster (70\,s vs.\ 3,000\,s on a 10,000\,x\,10,000 input) than the previous state of the art and exhibits sublinear scaling with input size. I also identify methods for using multidirectional flow routing and quickly eliminating landscape depressions and local minima. Tips for parallelization and a step-by-step guide to achieving it are given to help others achieve good performance with their own code. Complete, well-commented, easily adaptable source code for all versions of the algorithm is available as a supplement and on Github. 
  \end{abstract}

\end{frontmatter}

\section{Software Availability}
Complete, well-commented, easily-adaptable source code, an associated makefile, and correctness tests are available at \url{https://github.com/r-barnes/Barnes2018-Landscape}. The code is written in C++ using OpenACC for GPU acceleration and OpenMP for multi-core CPU acceleration. The code constitutes 3,304 lines of code spread across several implementations (averaging 367 lines of code per implementation) of which 42\% are or contain comments. %TODO

%This algorithm has been added to the RichDEM (\url{https://github.com/r-barnes/richdem}) terrain analysis suite, a collection of state of the art algorithms for processing large digital elevation models quickly.

\section{Introduction}

Models can be used to determine how landscapes are formed and to predict their future. However, doing so requires choosing between many possible governing equations~\citep{Tucker2010,Chen2014} and initial conditions. The analyses necessary to judge between the options require millions of model realizations and, to achieve statistical rigor, many thousands more.~\citep{Braun2013,Tucker2010} This computational cost is exacerbated by need for numerical stability and accuracy, which often requires using small time increments and high spatial resolutions. If is not feasible to perform these runs in serial. The algorithms I present below will enable landscape evolution researchers to achieve statistical rigor with minimal difficulty and computational cost.

Making use of SIMD, GPUs, and other accelerators will become increasingly important in the future. CPU clock-speeds are no longer increasing and, in some cases, are deliberately decreased to promote energy-efficiency. At the same time, core counts and data parallel architectures are becoming common. The algorithm and implementations I present here are better suited to this emerging paradigm than previous landscape evolution models.

The most efficient algorithm previously published is an $O(n)$ implicit integrator developed by \citet{Braun2013}. I will refer to this below as the B\&W algorithm and use it to benchmark the performance of the new code.

\begin{figure}
\centering
\includegraphics[width=\columnwidth]{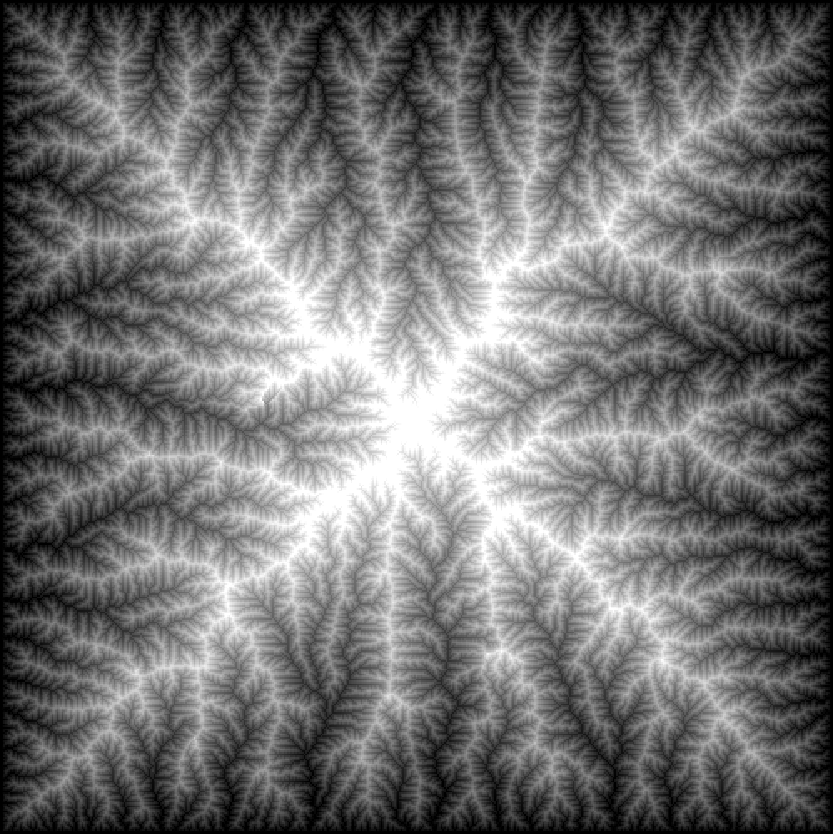}
\caption{An example output of the landscape evolution algorithm. \label{fig:output}}
\end{figure}

The design of the B\&W algorithm imposes serious limitations on parallelism and scalability; it is also limited to D8 flow routing. In contrast, my new algorithm can fully leverage SIMD (single instruction, multiple data) instructions such as the SSE and AVX families within a single core, distribute work without load imbalance between many cores, effectively offload work to accelerators such as GPUs, and work with multiple-flow direction routing. The algorithm produces outputs similar to that shown in \autoref{fig:output}.

\section{The Equation}

Though the algorithm described here could be applied to many equations governing the evolution of landscapes (\citep{Tucker2010,Chen2014} offer reviews), in this paper I use the stream power equation as an example as it has been widely used in the field.~\citep{Lague2014,Leigh2013,Whipple1999} In the equation, the evolution of the elevation $h$ of a point on a landscape is modeled as:
\begin{equation}
\label{equ:streampower}
\frac{\partial h}{\partial t}=-K A^{\hat m} \left(\frac{\partial h}{\partial x}\right)^{\hat n}
\end{equation}
where $K$ is a scalar influenced by lithology, channel width, and channel hydrology, among other possibilities; $A$ is the flow accumulation or contributing drainage area; and $\hat m$ and $\hat n$ are scaling constants. The implicit (backwards) Euler method using Newton-Raphson for the nonlinearity may be used to this equation.~\citep{Braun2013}.

Equation~\ref{equ:streampower} must be solved for all of the cells in an DEM. This requires a boundary condition (here I set the DEM's perimeter cells to a fixed base level, \citet{Braun2013} discusses other possibilities), the flow accumulation and slope at each point, and an ordering of points such that those receiving flow from a higher neighbour are processed before that neighbour. Below, I describe how to obtain these elements in a parallelized way.

\section{On Parallelism}
The following are a few notes on parallelism that guide algorithm design. The pseudocode in this paper instantiates these details at the conceptual level, but, since parallelism is maddeningly difficult~\citep{Lovecraft1928} to get right, the reader is advised to refer to the provided implementations for full details.

\subsection{Amdahl's law}
Amdahl's law says that a program's speed-up due to parallelism is bounded by the number of available parallel units and the time the program must spend running serial code; any serial code acts as a bottleneck. In B\&W only a subset of the steps of the algorithm are parallelized; therefore, as the number of parallel units increases, the run-time is dominated by the serial steps. Here, I overcome this by parallelizing all steps.

\subsection{Parallel For}
The algorithm consists of several distinct steps. Each step involves one or more loops over the elevation model, or portions thereof. These loops may be parallelized when their iterations are independent of each other. Such loops are denoted in the pseudocode with \textbf{for$_\parallel$}. For instance, during Uplift (\textsection\ref{sec:uplift}, Algorithm~\ref{alg:uplift}) each cell is raised by a constant factor. Since no cell needs information from any other cell for this to happen, all the cells may be uplifted concurrently.

\subsection{Concurrent Steps}
Sometimes, one or more steps may be executed concurrently. For example, the Flow Accumulation subroutine (\textsection\ref{sec:flow-accum}, Algorithm~\ref{alg:flow-accum}) does not depend on the elevations of the cells and the Uplift subroutine (\textsection\ref{sec:uplift}, Algorithm~\ref{alg:uplift}) does not depend on flow accumulation. As a result, the two steps could be run at the same time.

\subsection{Barriers and Synchronization}
Both OpenMP and OpenACC---two widely-used interfaces for parallel programming---require that all threads synchronize at the end of each parallel for region; this is known as an \textit{implicit barrier}. This prevents steps from being run concurrently. Eliminating barriers is vital to obtaining good parallel performance. In my implementations, I remove many implicit barriers, allowing threads to independently proceed through several steps before reaching a barrier. For simplicity the pseudocode does not show this, but readers can find full details in the reference implementation.

\subsection{Simplified Flow Control}
The presence of \textit{if} clauses within the inner loops of an algorithm can lead to slow downs (by a factor of 2 or more) when the CPU fails to predict which value the \textit{if} will take (failed branch prediction) or the GPU's warps diverge. The mere existence of an \textit{if} clause within a loop is often sufficient to prevent it from being vectorized for SIMD. To counter this, wherever possible, I try to keep the inner bodies of loops simple.

\section{The Algorithm}

\begin{table}
\centering
\begin{tabular}{lcccccccccc}
\hline
Cell   & 1 & 2 & 3 & 4 & 5 & 6 & 7 & 8 & 9 & 10 \\
Elev   & 3 & 2 & 3 & 4 & 1 & 2 & 3 & 2 & 4 & 3  \\
Rec    & 2 & 5 & 2 & 7 & X & 5 & 6 & 5 & 7 & 8  \\
Donor  & - & 1 & - & - & 2 & 7 & 4 & 10& - & -  \\
       & - & 3 & - & - & 6 & - & 9 & - & - & -  \\
       & - & - & - & - & 8 & - & - & - & - & -  \\
Dnum   & 0 & 2 & 0 & 0 & 3 & 1 & 2 & 1 & 0 & 0  \\
Queue  & 5 & 2 & 6 & 8 & 1 & 3 & 7 & 10& 4 & 9  \\
Levels & 0 & 1 & 4 & 8 & 10&   &   &   &   &    \\
Accum  & 1 & 3 & 1 & 1 & 10& 4 & 3 & 2 & 1 & 1  \\
\hline
\end{tabular}
\caption{Arrays used in the algorithm: a worked example. All arrays are zero-indexed. The entries of the \textbf{Cell} row refer to the node labels in Figure~\ref{fig:graphexplain}. \textbf{Elev}ations are chosen arbitrarily such that donor cells are higher than receiver cells. \textbf{Rec}eivers are calculated per Algorithm~\ref{alg:determine-receivers}. \textbf{Donor}s are calculated per Algorithm~\ref{alg:determine-donors}. Note that the Donor array should be read as snaking down one column, then down the next, and so on. Each column refers to one node's entries and each node has \textbf{Dmax} entries, some of which are unused (these are marked with dashes `-'). The \textbf{Dnum} array is the number of entries of each column of the Donor array which are filled in; that is, the number of Donors each cell has. The \textbf{Queue} array is the order in which cells should be processed, as determine by Algorithm~\ref{alg:generate-queue}; these values are the same as those shown in Figure~\ref{fig:stack-v-queue}b. The \textbf{Levels} array notes the 0-indexed beginnings of each level of parallelized cells, as marked by the dashed lines in Figure~\ref{fig:stack-v-queue}b. The flow \textbf{Accum}ulation array shows the flow accumulation of each cell, as determined by Algorithm~\ref{alg:flow-accum}. \label{tbl:example}}
\end{table}

\begin{figure}
\centering
\begin{tikzpicture}[scale=0.2,y=-1cm]
  \draw[fill=black] (8.01,17.82)  circle [radius=0.5] node (5)  [label={[xshift=0.4cm,yshift=-0.4cm]5}] {};
  \draw[fill=black] (4.09,10.44)  circle [radius=0.5] node (2)  [label={[xshift=0.35cm,yshift=-0.25cm]2}] {};
  \draw[fill=black] (9.28,11.68)  circle [radius=0.5] node (6)  [label={[xshift=0.3cm,yshift=-0.1cm]6}] {};
  %\draw             (9.28,11.68)  circle [radius=0.7] node (6)  [label={[xshift=0.35cm,yshift=-0.3cm]6}] {};
  \draw[fill=black] (14.53,11.01) circle [radius=0.5] node (8)  [label={[xshift=0.35cm,yshift=-0.4cm]8}] {};
  \draw[fill=black] (0.32,6.00)   circle [radius=0.5] node (1)  [label={[xshift=-0.3cm,yshift=-0.35cm]1}] {};
  \draw[fill=black] (5.89,5.75)   circle [radius=0.5] node (3)  [label={[xshift=0.3cm,yshift=-0.25cm]3}] {};
  %\draw             (9.63,7.13)   circle [radius=0.5] node (7)  [label={[xshift=0.35cm,yshift=-0.3cm]7}] {};
  \draw[fill=black] (9.63,7.13)   circle [radius=0.5] node (7)  [label={[xshift=0.35cm,yshift=-0.4cm]7}] {};
  \draw[fill=black] (18.03,7.44)  circle [radius=0.5] node (10) [label={[xshift=0.35cm,yshift=-0.3cm]10}] {};
  \draw[fill=black] (7.69,1.45)   circle [radius=0.5] node (4)  [label={[xshift=0cm,yshift=0cm]4}] {};
  \draw[fill=black] (14.82,4.02)  circle [radius=0.5] node (9)  [label={[xshift=0cm,yshift=0cm]9}] {};

  \draw[-{Latex[length=2.5mm, width=2mm]}] (2)  -- (5);
  \draw[-{Latex[length=2.5mm, width=2mm]}] (6)  -- (5);
  \draw[-{Latex[length=2.5mm, width=2mm]}] (8)  -- (5);
  \draw[-{Latex[length=2.5mm, width=2mm]}] (1)  -- (2);
  \draw[-{Latex[length=2.5mm, width=2mm]}] (3)  -- (2);
  \draw[dashed] (3)  -- (6);
  \draw[-{Latex[length=2.5mm, width=2mm]}] (7)  -- (6);
  \draw[dashed] (7)  -- (8);
  \draw[-{Latex[length=2.5mm, width=2mm]}] (10) -- (8);
  \draw[dashed] (4) -- (1);
  \draw[dashed] (4) -- (3);
  \draw[-{Latex[length=2.5mm, width=2mm]}] (4) -- (7);
  \draw[-{Latex[length=2.5mm, width=2mm]}] (9) -- (7);
  \draw[dashed] (9) -- (10);
  \draw[dashed] (4) -- (9);
  \draw[dashed] (9) -- (8);
  \draw[dashed] (3) -- (7);
  \draw[dashed] (1) -- (3);
  \draw[dashed] (2) -- (6);
  \draw[dashed] (6) -- (8);

  \draw[line width=2] (7,17.82) -- (9,17.82);

  %\draw[line width=4] ([xshift=1cm]5) -- ([xshift=-1cm]5);
\end{tikzpicture}
\caption{Elevation nodes and their connections. Solid arrows denote flow along the path of greatest slope while dashed lines denoted possible flow routes of lesser slope which are modeled as having no flow. In this example, Nodes 4 and 9 are the \textbf{donors} of Node 7 and Node 6 is the \textbf{receiver} of Node 7. Node 5 is at the base level (marked by the solid line) and its elevation does not change. Figure adapted from \citep{Braun2013}.\label{fig:graphexplain}}
\end{figure}
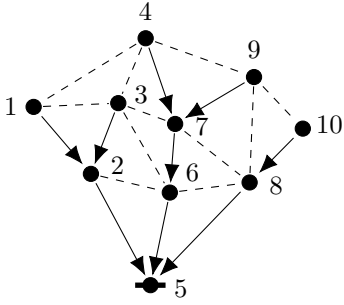

The algorithm models each grid cell as having receiver nodes (those receiving flow from an upslope neighbour) and donor nodes (those nodes which pass their flow to a downslope neighbour). Figure~\ref{fig:graphexplain} depicts these concepts.

The algorithm assumes that a regular, 4-, 6-, or 8-connected grid is used. This is important since it provides a simple addressing mechanism for cells and enables the processor to intelligently prefetch data from RAM, which is slow, and maintain it in the L1, L2, and L3 caches, which are fast.~\citep{Drepper2007,Stark2011} It also permits efficient transfer of memory between the CPU and GPU.

Table~\ref{tbl:example} shows a worked example of the arrays developed in the following algorithms.

\subsection{Step 1: Initialization}
The algorithm requires several global variables. These are as follows:
\begin{itemize}
\item \textit{Dmax}:             The maximum number of potential donors of any cell in the elevation model. For a rectangular grid with horizontal, vertical, and diagonal connections, this is 8.
\item \textit{$\hat m$}:         The exponent of the flow accumulation area in the stream power equation (Equ.~\ref{equ:streampower}).
\item \textit{$\hat n$}:         The exponent of the local slope in the stream power equation (Equ.~\ref{equ:streampower}).
\item \textit{$\hat u$}:         The rate of uplift.
\item \textsc{NoFlow}:           A constant indicating that the cell has no receiver.
\item \textit{$\epsilon$}:       The tolerance for convergence in the Newton-Raphson method.
% \item \textit{$T_\textrm{max}$}: The maximum number of parallel threads used.
% \item \textit{$T_\textrm{N}$}:   The number of the current thread.
% \item \textit{$W_\textrm{Q}$}:   The number of queue entries each thread gets.
% \item \textit{$W_\textrm{L}$}:   The number of level entries each thread gets.
\end{itemize}

The algorithm requires one input array:
\begin{itemize}
\item \textbf{Elev}: The height/elevation model. This is a one-dimensional array of size \textit{width} by \textit{height}. A particular cell at location $(x,y)$ is addressed as $y\cdot\textit{width}+x$.
\end{itemize}

\subsection{Step 2: Determine Receivers}
Here, for each cell $c$, we determine which of $c$'s neighbours receives its flow, choosing the neighbour with the greatest downhill slope. The address of the receiving neighbour is stored in the \textit{Rec} array. Each entry in this array has a corresponding cell in the \textit{Elev} array. Note that cells on the perimeter of the model do not transfer flow.

\begin{breakablealgorithm}
\caption{\small {\sc Determine Receivers}}
\label{alg:determine-receivers}
\footnotesize
\begin{algorithmic}[1]
\State Let $Rec$ have the same dimensions as $Elev$
\State Initialize $Rec$ to \textsc{NoFlow}
\State
\MultiForAll{cells $c$ in the interior of $Elev$}
  \State $s_\textrm{max}\gets0$ \Comment Maximum slope
  \State $n_\textrm{max}\gets\textsc{NoFlow}$ \Comment Neighbour with that slope
  \ForAll{neighbours $n$ of $c$}
    \State $s\gets (Elev[c]-Elev[n])/\textrm{dist}(c,n)$
    \State \Comment Slope from $c$ to $n$
    \If{$s>s_\textrm{max}$}
      \State $s_\textrm{max}\gets s$
      \State $n_\textrm{max}\gets n$
    \EndIf
  \EndFor
  \State $Rec[c]\gets n_\textrm{max}$
\EndMultiForAll
\end{algorithmic}
\end{breakablealgorithm}

\subsection{Step 3: Determine Donors}
The \textit{Donors} array is an inversion of the \textit{Rec} array. Each cell in \textit{Elev} corresponds to \textit{Dmax} entries in this array, where each entry denotes the address of a cell from which flow is received. Thus, the address of the cells from which a particular cell $(x,y)$ will receive flow are given by $\textit{Dmax}\cdot(y\cdot\textit{width}+x)+k=\textit{Dmax}\cdot c+k$ for $k\in[0,\textit{Dnum}(c))$, where $\textit{Dnum}(c)$ indicates the number of neighbours from which $c$ receives flow.

In the B\&W algorithm, each donating cell informs its receiver that it will be receiving a donation. This prevents parallelization because multiple donor cells may pass their information at the same time: a race condition. This could be prevented with atomic operations, but a more performant solution is to have each cell identify its donors.

\begin{breakablealgorithm}
\caption{\small {\sc Determine donors}}
\label{alg:determine-donors}
\footnotesize
\begin{algorithmic}[1]
\State Let $Donor$ have the same dimensions as $Elev\cdot\textit{Dmax}$
\State Let $\textit{Dnum}$ have the same dimensions as $Elev$
\State
\MultiForAll{cells $c$ in the interior of $Elev$} \label{alg:donorsloop}
  \State $\textit{Dnum}[c]\gets 0$
  \ForAll{neighbours $n$ of $c$}
    \If{$\textit{Rec}[n]=c$}
      \State $Donor[\textit{Dmax}\cdot c+\textit{Dnum}[c]]\gets n$
      \State $\textit{Dnum}[c]\gets \textit{Dnum}[c]+1$
    \EndIf
  \EndFor
\EndMultiForAll
\end{algorithmic}
\end{breakablealgorithm}

\subsection{Step 4: Generate Queue}
\label{sec:queue}

\begin{figure}
\centering
 
\begin{subfigure}[t]{0.48\columnwidth}
    \centering
\begin{tikzpicture}[scale=0.15,y=-1cm]
  \draw[fill=black] (8.01,17.82)  circle [radius=0.5] node (5)  [label={[xshift=0.4cm,yshift=-0.4cm]1}] {};
  \draw[fill=black] (4.09,10.44)  circle [radius=0.5] node (2)  [label={[xshift=0.3cm,yshift=-0.35cm]2}] {};
  \draw[fill=black] (9.28,11.68)  circle [radius=0.5] node (6)  [label={[xshift=0.25cm,yshift=-0.35cm]5}] {};
  \draw[fill=black] (14.53,11.01) circle [radius=0.5] node (8)  [label={[xshift=0.25cm,yshift=-0.4cm]9}] {};
  \draw[fill=black] (0.32,6.00)   circle [radius=0.5] node (1)  [label={[xshift=-0.3cm,yshift=-0.35cm]3}] {};
  \draw[fill=black] (5.89,5.75)   circle [radius=0.5] node (3)  [label={[xshift=0.2cm,yshift=-0.25cm]4}] {};
  \draw[fill=black] (9.63,7.13)   circle [radius=0.5] node (7)  [label={[xshift=0.25cm,yshift=-0.45cm]6}] {};
  \draw[fill=black] (18.03,7.44)  circle [radius=0.5] node (10) [label={[xshift=0.35cm,yshift=-0.3cm]10}] {};
  \draw[fill=black] (7.69,1.45)   circle [radius=0.5] node (4)  [label={[xshift=0cm,yshift=0cm]7}] {};
  \draw[fill=black] (14.82,4.02)  circle [radius=0.5] node (9)  [label={[xshift=0cm,yshift=0cm]8}] {};

  \draw[-{Latex[length=2.5mm, width=2mm]}] (2)  -- (5);
  \draw[-{Latex[length=2.5mm, width=2mm]}] (6)  -- (5);
  \draw[-{Latex[length=2.5mm, width=2mm]}] (8)  -- (5);
  \draw[-{Latex[length=2.5mm, width=2mm]}] (1)  -- (2);
  \draw[-{Latex[length=2.5mm, width=2mm]}] (3)  -- (2);
  \draw[-{Latex[length=2.5mm, width=2mm]}] (7)  -- (6);
  \draw[-{Latex[length=2.5mm, width=2mm]}] (10) -- (8);
  \draw[-{Latex[length=2.5mm, width=2mm]}] (4) -- (7);
  \draw[-{Latex[length=2.5mm, width=2mm]}] (9) -- (7);

  \draw[line width=2] (7,17.82) -- (9,17.82);
\end{tikzpicture}
    \caption{Depth-First/Stack Order}
\end{subfigure}
~
\begin{subfigure}[t]{0.48\columnwidth}
    \centering
\begin{tikzpicture}[scale=0.15,y=-1cm]
  \draw[fill=black] (8.01,17.82)  circle [radius=0.5] node (5)  [label={[xshift=0.4cm,yshift=-0.4cm]1}] {};
  \draw[fill=black] (4.09,10.44)  circle [radius=0.5] node (2)  [label={[xshift=0.3cm,yshift=-0.35cm]2}] {};
  \draw[fill=black] (9.28,11.68)  circle [radius=0.5] node (6)  [label={[xshift=0.25cm,yshift=-0.35cm]3}] {};
  \draw[fill=black] (14.53,11.01) circle [radius=0.5] node (8)  [label={[xshift=0cm,yshift=-0.6cm]4}] {};
  \draw[fill=black] (0.32,6.00)   circle [radius=0.5] node (1)  [label={[xshift=-0.3cm,yshift=-0.35cm]5}] {};
  \draw[fill=black] (5.89,5.75)   circle [radius=0.5] node (3)  [label={[xshift=0.2cm,yshift=-0.25cm]6}] {};
  \draw[fill=black] (9.63,7.13)   circle [radius=0.5] node (7)  [label={[xshift=0.35cm,yshift=-0.4cm]7}] {};
  \draw[fill=black] (18.03,7.44)  circle [radius=0.5] node (10) [label={[xshift=0.35cm,yshift=-0.3cm]8}] {};
  \draw[fill=black] (7.69,1.45)   circle [radius=0.5] node (4)  [label={[xshift=0cm,yshift=0cm]9}] {};
  \draw[fill=black] (14.82,4.02)  circle [radius=0.5] node (9)  [label={[xshift=0cm,yshift=0cm]10}] {};

  \draw[-{Latex[length=2.5mm, width=2mm]}] (2)  -- (5);
  \draw[-{Latex[length=2.5mm, width=2mm]}] (6)  -- (5);
  \draw[-{Latex[length=2.5mm, width=2mm]}] (8)  -- (5);
  \draw[-{Latex[length=2.5mm, width=2mm]}] (1)  -- (2);
  \draw[-{Latex[length=2.5mm, width=2mm]}] (3)  -- (2);
  \draw[-{Latex[length=2.5mm, width=2mm]}] (7)  -- (6);
  \draw[-{Latex[length=2.5mm, width=2mm]}] (10) -- (8);
  \draw[-{Latex[length=2.5mm, width=2mm]}] (4) -- (7);
  \draw[-{Latex[length=2.5mm, width=2mm]}] (9) -- (7);

  \draw[line width=2] (7,17.82) -- (9,17.82);

  \centerarc[black,dashed](5)(200:350:4);
  \centerarc[black,dashed](5)(220:330:10);
  \centerarc[black,dashed](5)(237:320:15);
  % Syntax: [draw options] (center) (initial angle:final angle:radius)
\end{tikzpicture}
    \caption{Breadth-First/Queue Order}
\end{subfigure}

\caption{Stack vs.\ queue: illustrated. The B\&W algorithm uses a stack, shown in \textbf{(a)}, to order cells for processing. This results in a depth-first traversal in which the first of a node's children is visited, and then the first of that node's children, and so on. When there are no more children, the algorithm back-tracks one level and processes the next child, if there is one. This child is on a different branch. Using a queue, shown in \textbf{(b)}, results in a breadth-first traversal. Nodes are visited in an expanding wave (shown as dashed lines) from the source. Note that in the stack, 2 is processed and then 3: this operation cannot be parallelized. To work around this, 2 and its children could be processed in parallel with 5 and its children, but note that the subtrees are of different sizes. Note also, that Node 1 has only three children, limiting the amount of parallelism to three threads. In contrast, the queue makes it easy to identify that 2--4 and 5--8 can be processed in parallel, and that there is a greater opportunity for paralleism (4 rather than 3). Figure~\ref{fig:stack-queue-square} expands on this. \label{fig:stack-v-queue}}
\end{figure}
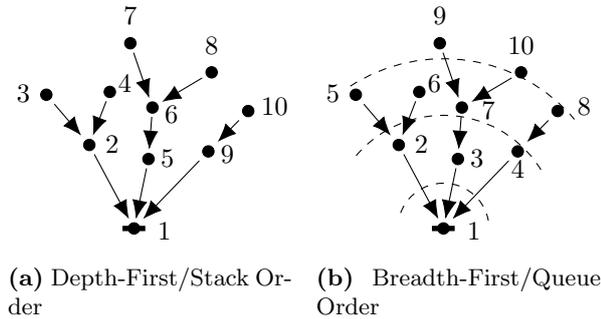

% g++ fastscape_serial1.cpp  Timer.cpp ; ./a.out | grep '\\' > paper/fig_stack.tex
% g++ fastscape_openacc3.cpp Timer.cpp ; ./a.out | grep '\\' > paper/fig_queue.tex
\begin{figure}
\centering

{\large Cells That Can Be Processed In Parallel}

\begin{subfigure}[t]{0.48\columnwidth}
  \centering
  \footnotesize Cells of the same colour must be run sequentially

  \includegraphics[width=1\columnwidth]{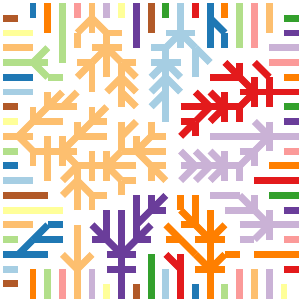}

  \caption{Stack Order}
\end{subfigure}
~
\begin{subfigure}[t]{0.48\columnwidth}
  \centering
  \footnotesize Cells of the same colour may be run in parallel

  \includegraphics[width=1\columnwidth]{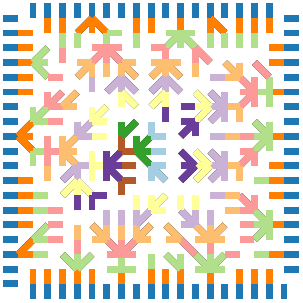}

  \caption{Queue Order}
\end{subfigure}

\vspace{0.5cm}

{\large Which Thread Processes Which Cells}

\footnotesize Cells of the same colour are processed by the same thread

\begin{subfigure}[t]{0.48\columnwidth}
  \centering

  \includegraphics[width=1\columnwidth]{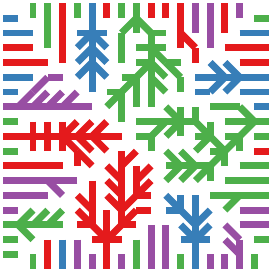}
  \caption{Stack Order}
\end{subfigure}
~
\begin{subfigure}[t]{0.48\columnwidth}
  \centering

  \includegraphics[width=1\columnwidth]{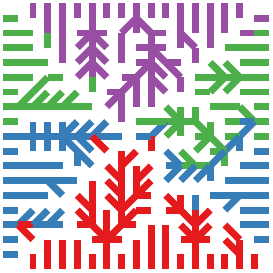}
  \caption{Queue Order}
\end{subfigure}

\vspace{0.5cm}

{\large When Cells Are Processed}

\footnotesize Redder cells are processed later

\begin{subfigure}[t]{0.48\columnwidth}
  \centering

  \includegraphics[width=1\columnwidth]{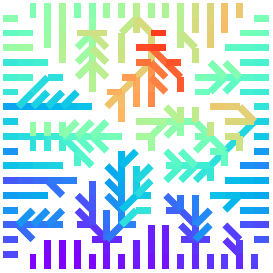}
  \caption{Stack Order}
\end{subfigure}
~
\begin{subfigure}[t]{0.48\columnwidth}
  \centering

  \includegraphics[width=1\columnwidth]{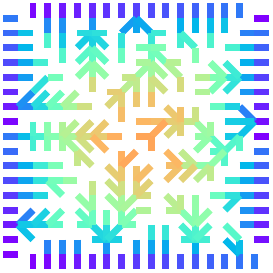}
  \caption{Queue Order}
\end{subfigure}

\caption{Stack vs.\ queue: illustrated on a larger example using four threads. Note that using a stack/depth-first traversal leads to many small trees and a few large ones. This causes load imbalance and makes this form of the algorithm challenging to parallelize. In contrast, the queue can parallelize many cells at each level. The threads process 70, 71, 74, and 74 cells each in the queue example and 48, 51, 85, and 105 cells each in the stack example. Figure~\ref{fig:stack-v-queue} shows a smaller example for which the algorithm is explained. \label{fig:stack-queue-square}}
\end{figure}

% rick@sally:~/projects/quickscape/paper/tree_fig_draw$ cat fig_queue_inc.tex | sed 's/.*=//' | sed 's/\].*//' | sort | uniq -c
%      74 tcolor0
%      74 tcolor1
%      71 tcolor2
%      70 tcolor3
% rick@sally:~/projects/quickscape/paper/tree_fig_draw$ cat fig_stack_inc.tex | sed 's/.*=//' | sed 's/\].*//' | sort | uniq -c
%      85 tcolor0
%      51 tcolor1
%     105 tcolor2
%      48 tcolor3

The \textit{Queue} array stores the addresses of cells in the order they are to be processed. Processing the cells in this order ensures that the information needed to solve the stream power equation is always available. Each cell appears in this array once. The levels array contains indices corresponding to subdivisions of \textit{Queue}. The cells in each subdivision may be processed concurrently.

%auto tstack    = &stack[tnum*t_stack_width];
%auto tlevels   = &levels[tnum*t_level_width];
%int &tnlevel   = nlevel[tnum];

Note that at this stage the algorithm differs from the B\&W variant in a fundamental way: B\&W uses a stack whereas I use a queue. From the perspective of graphs this is the difference between depth-first and breadth-first traversal, respectively. The difference is illustrated in Figure~\ref{fig:stack-v-queue} and, again, in Figure~\ref{fig:stack-queue-square}. As explained in \textsection\ref{sec:breadth}, this greatly increases potential parallelism.

To build \textit{Queue}, all of the cells without receivers (the mouths of rivers and pits of depressions) are first added to the queue. A note is made in \textit{Levels} of how many of these cells there are. Next, all of these cells' donors are added and another note is made in \textit{Levels}. And then the donors of the donors are added, and so on.

Parallelizing this step is difficult. It is written here as a serial algorithm. Parallelism strategies are discussed below in \textsection\ref{sec:resultdisc}.

\begin{breakablealgorithm}
\caption{\small {\sc Generate Queue}}
\label{alg:generate-queue}
\footnotesize
\begin{algorithmic}[1]
\State Let \textit{Levels} be a vector that is ``sufficiently long"
\State $Levels[0]\gets1$
\State $L_s\gets 1$  \Comment Write location in \textit{Levels}
\State $\textit{nqueue}\gets 0$  \Comment Open location in \textit{Queue}
\State
\ForAll{cells $c$ in $Rec$} \label{alg:queueloop}
  \If{$Rec[c]=\textsc{NoFlow}$} \Comment Is it a source cell?
    \State $\textit{Queue}[\textit{nqueue}]\gets c$
    \State $\textit{nqueue}\gets\textit{nqueue} + 1$
  \EndIf
\EndFor
\State $\textit{Levels}[L_s]=\textit{nqueue}$ \Comment Note last cell of \textit{Levels}
\State $L_s\gets L_s + 1$ 
\State
\State $L_L\gets -1$  \Comment Lower index of current level
\State $L_U\gets  0$  \Comment Upper index of current level
\While{$L_L<L_U$}
  \State $L_L\gets L_U$
  \State $L_U\gets\textit{nqueue}$
  \For{$c\in [L_L,LU)$}
    \ForAll{$k\in \textit{Dnum}[c]$}
      \State $\textit{Queue}[\textit{nqueue}]\gets \textit{donor}[\textit{Dmax}\cdot c+k]$
      \State $\textit{nqueue}\gets \textit{nqueue} + 1$
    \EndFor
  \EndFor
  \State $\textit{Levels}[L_s]\gets\textit{nstack}$
  \State $L_s\gets L_s+1$
\EndWhile
\State $L_s\gets L_s-1$ \Comment Correct overshoot
\end{algorithmic}
\end{breakablealgorithm}

\subsection{Step 5: Compute Flow Accumulation}
\label{sec:flow-accum}
The \textit{Accum} array stores the flow accumulation (also known as drainage area, contributing area, and upslope area) of each cell. As described by \citet{OCallaghan1984} and \citet{Mark1987}, the flow accumulation $A$ of a cell $c$ is defined recursively as
\begin{equation}
\label{equ:flow_acc}
A(c)=w(c)+\sum_{n\in \mathcal{N}(c)} \alpha(n,c) A(n)
\end{equation}
where $w(c)$ is the amount of flow which originates at the cell $c$; frequently, this is taken to be 1, but the value can also vary across a DEM if, for example, rainfall or soil absorption differs spatially. The summation is across all of the cell $c$'s neighbours $\mathcal{N}(c)$. $\alpha(n,c)$ represents the fraction of the neighbouring cell's flow accumulation $A(c)$ which is apportioned to $c$. Flow may be absorbed during its downhill movement, but may only be increased by cells, so $\alpha$ is constrained such that for a given cell $c$, $\sum_n \alpha(c,n)\le1$. Flow accumulation may be parallelized across each level of the queue (see Fig.~\ref{fig:stack-queue-square}b).

\begin{breakablealgorithm}
\caption{\small {\sc Flow Accumulation}}
\label{alg:flow-accum}
\footnotesize
\begin{algorithmic}[1]
\State Let $A$ have the same dimensions as $Elev$
\State Initialize $A$ to $\Delta x \Delta y$
\State
\For{$l\in[L_s-2,0]$}
  \MultiFor{$c\in[\textit{Levels}[l],\textit{Levels}[l+1])$}
    \For{$i\in[0,\textit{Dnum}[c])$}
      \State $n\gets Donor[\textit{Dmax}\cdot c+i]$
      \State $A[c]\gets A[c] + A[n]$
    \EndFor
  \EndMultiFor  
\EndFor
\end{algorithmic}
\end{breakablealgorithm}

\subsection{Step 6: Uplift}
\label{sec:uplift}
Tectonic uplift is incorporated in a straight-forward manner: every cell is elevated at some rate $\hat{u}$. (Note that parallelism is still trivial if uplift varies spatially.)
\begin{breakablealgorithm}
\caption{\small {\sc Uplift}}
\label{alg:uplift}
\footnotesize
\begin{algorithmic}[1]
\MultiForAll{cells $c$ in the interior of $Elev$}
  \State $Elev[c]\gets Elev[c]+\hat{u} \Delta t$
\EndMultiForAll
\end{algorithmic}
\end{breakablealgorithm}

\subsection{Step 7: Calculate Erosion}
Finally, erosion is calculated by implementing the Newton-Raphson method~\citep{Braun2013}. Note that the cells within each level are neither receivers nor donors of each other. More importantly, there is no causal connection between them. This means that all of the cells in a level can be executed in parallel, as in Algorithm~\ref{alg:erosion}, Line~\ref{alg:erosionloop}. Note that the tolerance check on Line~\ref{algline:tolcheck} could be replaced with a fixed number of loops if the maximum number required were known.

\begin{breakablealgorithm}
\caption{\small {\sc Calculate erosion}}
\label{alg:erosion}
\footnotesize
\begin{algorithmic}[1]
\For{$l\in[1,L_s)$}
  %Parallel
  \MultiFor{$i\in[\textit{Levels}[l],\textit{Levels}[l+1])$} \label{alg:erosionloop}
    \State $c\gets \textit{Queue}[i]$ \Comment Current focal cell
    \State $n\gets Rec[c]$            \Comment Neighbour of focal cell
    \State $F\gets K\cdot \Delta t \cdot \textit{Acc}[c]^m / \textrm{dist}(n,c)^n$
    \State $h_0 \gets \textit{Elev}[c]$ \Comment Elevation of focal cell
    \State $h_n \gets \textit{Elev}[n]$ \Comment Elevation of neighbour cell
    \State $h_\textrm{new} \gets h_0$   \Comment Current updated value of $Elev[c]$
    \State $h_p \gets h_0$              \Comment Previous updated value of $Elev[c]$
    \State $\delta \gets 2\epsilon$     \Comment Difference between updated values
    \While{$|\delta|>\epsilon$} \label{algline:tolcheck}  \Comment While difference $>$ tolerance
      \State $h_\textrm{new} \gets h_\textrm{new}-(h_\textrm{new}-h_0+F\cdot (h_\textrm{new}-h_n)^n)/(1+F \cdot n\cdot (h_\textrm{new}-h_n)^{n-1})$
      \State $\delta \gets h_\textrm{new}-h_p$  \Comment Update difference
      \State $h_p \gets h_\textrm{new}$         \Comment New previous value
    \EndWhile
    \State $Elev[c]\gets h_\textrm{new}$   \Comment New elevation for focal cell
  \EndMultiFor
\EndFor
\end{algorithmic}
\end{breakablealgorithm}

\subsection{Rinse, Repeat}
All of the above steps, excluding initialization, are repeated as many times as necessary until the desired interval of time has been simulated.

\section{The Improvements, Explained}
\subsection{Breadth-first Traversal}
\label{sec:breadth}

% \begin{figure}
% % g++ fastscape_serial1.cpp  Timer.cpp ; ./a.out | grep '\\' > paper/fig_stack.tex
% % g++ fastscape_openacc3.cpp Timer.cpp ; ./a.out | grep '\\' > paper/fig_queue.tex
% \centering
% \begin{subfigure}{\columnwidth}
%   \centering
%   \includegraphics[width=\columnwidth]{para_short.pdf}
%   %\caption{}
% \end{subfigure}

% \begin{subfigure}{\columnwidth}
%   \centering
%   \includegraphics[width=\columnwidth]{para_long.pdf}
%   %\caption{}
% \end{subfigure}

% \caption{Available parallelism vs.\ iteration. The top graph and bottom graphs show different views of the same data. Note that the queue utilized by the new variant has high parallelism and a low number of iterations in contrast to the stack employed by the B\&W algorithm, which has low parallelism and a high number of iterations. \label{fig:avail-parallelism}} %Actually, 501x501
% \end{figure}

As described above, the this new variant of the B\&W algorithm utilizes a breadth-first traversal rather than the depth-first traversal used in the original algorithm. That is: level sets are formed, rather than rooted trees. Figure~\ref{fig:stack-v-queue} depicts the difference between these on a small sample drawn from \citet{Braun2013}; however, focusing on this small sample obscures important meta-level properties of the algorithms which are only visible on larger datasets such as that shown in Figure~\ref{fig:stack-queue-square}.

As Figure~\ref{fig:stack-queue-square} shows, processing stacks in parallel, as suggested by \citep{Braun2013}, initially results in a high degree of parallelism (equal to the number of edge cells of the elevation model). However, many of the stacks are small. As a result, much of the available parallelism is quickly exhausted until a single thread is operating on a single, usually large, stack. This is known as load imbalance, and is a serious problem in parallel computing.

One way to overcome this is to launch a new parallel task every time a tree branches; eventually, all of the available parallelism will be used. However, this is not a good solution: there is a significant overhead, on the order of microseconds, to starting OpenMP tasks.~\citep{Bull2012} Since each task would process a single node, the overhead of starting a task is likely to exceed the work done by that task. Another potential solution is to only launch tasks when large trees branch. But this begs the question of how big a tree should be and how long it would take to determine which trees to branch.

In contrast, a breadth-first traversal provides an easy route to guaranteed parallelism. Since each cell donates to at most one receiver, the donors of the base level cells can all be processed in parallel, as can the donors of the donors, and so on. At each level the set of independent cells is exactly known and easily identified.

%To demonstrate the difference, 2,400 random 501\,x\,501 cell elevation models were constructed and evolved for 120 timesteps. At this point the number of trees in the stack and the number of cells in the queue were counted for each iteration of Algorithm~\ref{alg:erosion}. The results are shown in Figure~\ref{fig:avail-parallelism}. The new algorithm shows sustained parallel potential of hundreds to thousands of cells and a low maximum number of iterations. In contrast, the B\&W algorithm has low parallel potential and a high maximum number of iterations.

Parallelism can be realized in one of several ways. First, on a single core, SIMD (single-instruction, multiple-data) instructions can be used. These allow the same operation to be applied to several elements of an array at once. The latest such instruction set, AVX-512, can process 16 single-precision or 8 double-precision values at once. The B\&W algorithm cannot take advantage of SIMD since each thread operates on a separate tree and each tree is inherently sequential.

Second, OpenMP may be used to easily divide an array between separate threads/cores. This permits the full power of a CPU to be used. For example, the new Summit supercomputer at Oak Ridge National Lab has 96 SIMD units per socket, allowing for up to 1,536 single-precision cells to be processed at once. In contrast, each socket has only 24 cores, which is the maximum parallelism that could be applied to the original B\&W algorithm.

Third, GPUs provide an avenue to even greater parallelism. The Nvidia Volta GPUs used by Summit allow for approximately 5120 simultaneous floating-point operations. Each node has several GPUs.

Thus, the design of the B\&W algorithm limits parallelism to tens of cells at a time, whereas the new design presented here permits many thousands of simultaneous operations.

\subsection{Local Minima}
It is often desirable to calculate flow directions only after internally-draining regions of a digital elevation model such as depressions and pits (see \citet{Lindsay2015} for a typology) have been eliminated. This ensures that all flows reach the edge of the model. Depressions may arise spuriously from random initial conditions, or may also represent endorheic basins. Regardless, they are usually a transient feature.

Depressions may be dealt with in one of three ways.
\begin{itemize}
\item They can be ignored. Over time, the model's erosive processes will either fill them or create outlets.
\item The depressions can be filled to the level of their lowest outlets. This is the method recommended by \citet{Braun2013}, who suggest an $O(N\sqrt{N})$ algorithm. This is suboptimal. Optimal theoretical and empirical performance is achieved by the Priority-Flood algorithm identified in a review by \citet{Barnes2014pf}. On integer (or appropriately discretized floating-point) data Priority-Flood runs in $O(N)$ time. For general floating-point data, it runs in $O(m \log m)$ time where $m\ll N$. Recent work by \citet{Zhou2016} and \citet{Wei2018} has led to significant reductions in run-time. For larger models, \citet{Barnes2016} presents an optimal parallelization of Priority-Flood.
\item Depressions which are small or shallow may be breached by cutting a channel from a depression's pit to some point beyond its outlet, as detailed by \citet{Lindsay2015}.
\end{itemize}

The filling of depressions may result in flat regions where there is no locally-defined flow direction. If desired, such regions may be resolved either (a)~as part of Priority-Flood~\citep{Barnes2014pf} or (b)~by routing flow both away from higher terrain and towards lower terrain~\citep{Barnes2014dd}.

\subsection{Larger Models}
For truly large elevation models, \citet{Barnes2016} and \citet{Barnes2017} describe optimal parallel algorithms for performing depression-filling and flow accumulation. These algorithms can efficiently process trillions of cells. Although such datasets are presently larger than those used in the context of landscape evolution, they may be of interest in the future.

\section{Multiple Flow Directions}

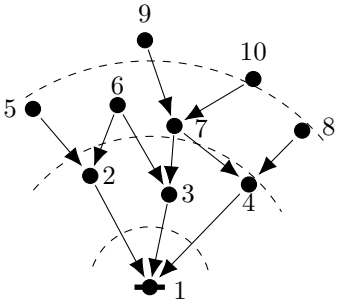
\begin{figure}
\centering
\begin{tikzpicture}[scale=0.20,y=-1cm]
  \draw[fill=black] (8.01,17.82)  circle [radius=0.5] node (5)  [label={[xshift=0.4cm,yshift=-0.4cm]1}] {};
  \draw[fill=black] (4.09,10.44)  circle [radius=0.5] node (2)  [label={[xshift=0.25cm,yshift=-0.35cm]2}] {};
  \draw[fill=black] (9.28,11.68)  circle [radius=0.5] node (6)  [label={[xshift=0.25cm,yshift=-0.35cm]3}] {};
  \draw[fill=black] (14.53,11.01) circle [radius=0.5] node (8)  [label={[xshift=0.0cm,yshift=-0.6cm]4}] {};
  \draw[fill=black] (0.32,6.00)   circle [radius=0.5] node (1)  [label={[xshift=-0.3cm,yshift=-0.35cm]5}] {};
  \draw[fill=black] (5.89,5.75)   circle [radius=0.5] node (3)  [label={[xshift=0.0cm,yshift=-0.1cm]6}] {};
  \draw[fill=black] (9.63,7.13)   circle [radius=0.5] node (7)  [label={[xshift=0.35cm,yshift=-0.4cm]7}] {};
  \draw[fill=black] (18.03,7.44)  circle [radius=0.5] node (10) [label={[xshift=0.35cm,yshift=-0.3cm]8}] {};
  \draw[fill=black] (7.69,1.45)   circle [radius=0.5] node (4)  [label={[xshift=0cm,yshift=0cm]9}] {};
  \draw[fill=black] (14.82,4.02)  circle [radius=0.5] node (9)  [label={[xshift=0cm,yshift=0cm]10}] {};

  \draw[-{Latex[length=2.5mm, width=2mm]}] (2)  -- (5);
  \draw[-{Latex[length=2.5mm, width=2mm]}] (6)  -- (5);
  \draw[-{Latex[length=2.5mm, width=2mm]}] (8)  -- (5);
  \draw[-{Latex[length=2.5mm, width=2mm]}] (1)  -- (2);
  \draw[-{Latex[length=2.5mm, width=2mm]}] (3)  -- (2);
  \draw[-{Latex[length=2.5mm, width=2mm]}] (7)  -- (6);
  \draw[-{Latex[length=2.5mm, width=2mm]}] (10) -- (8);
  \draw[-{Latex[length=2.5mm, width=2mm]}] (4) -- (7);
  \draw[-{Latex[length=2.5mm, width=2mm]}] (9) -- (7);
  \draw[-{Latex[length=2.5mm, width=2mm]}] (3)  -- (6); %MFD arrow
  \draw[-{Latex[length=2.5mm, width=2mm]}] (7)  -- (8); %MFD arrow  
  %\draw[dashed,-{Latex[length=2mm, width=1mm]}] (3)  -- (7); %MFD arrow  

  \draw[line width=2] (7,17.82) -- (9,17.82);

  \centerarc[black,dashed](5)(200:350:4);
  \centerarc[black,dashed](5)(220:330:10);
  \centerarc[black,dashed](5)(237:320:15);
  % Syntax: [draw options] (center) (initial angle:final angle:radius)
\end{tikzpicture}

\caption{Multiple-flow directions. When multiple-flow directions are present a cell may have multiple receivers. All of a focal cell's receivers must be processed before the focal cell can be processed. In the figure above it is clear that this necessitates the queue order used in the new algorithm. Cells are numbered in the order they should be processed.
%Note that the dashed arrow from 6 to 7 would break the queue's wavefront.
\label{fig:mfd}}
\end{figure}

The B\&W uses the D8 flow router~\citep{OCallaghan1984,Mark1987}. This models flow as descending along the path of steepest slope from a cell to a single one of its neighbours (provided there is a local gradient). This implies the convenient property that flows only converge and never diverge. As a result, each cell has only a single receiver and Equation~\ref{equ:streampower} is solved with respect to only a single pair of cells: one upslope, the other down. Multiple-flow direction (MFD) routers \citep{Quinn1991,Freeman1991,Holmgren1994,Pilesjo1998,Seibert2007,Tarboton1997,Orlandini2003,Orlandini2009} break this assumption. As Figure~\ref{fig:mfd} shows, the queue used by the new algorithm is the right choice for working with MFD. %Though there are limitations. In Figure~\ref{fig:mfd}, if cell 6 connected to cell 7 along the dashed arrow, then the wavefront would be broken and a queue would not be adequate

\begin{comment}
Equation~\ref{equ:streampowerexpanded} must be modified for MFD:
\begin{equation}
\label{equ:streampowermfd}
\frac{\partial h_c}{\partial t}=-\sum_{i \in \mathcal{N}} K A_i^{\hat m} \left(\frac{\partial h_c}{\partial x_i}\right)^{\hat n}
\end{equation}
where $h_c$ is the elevation of the focal cell, $i$ is a particular neighbour from its set of neighbours $\mathcal{N}$, $A_i$ is the flow accumulation apportioned to $i$, and $\partial h_c/\partial x_i$ approximates the slope between $c$ and $i$.
\end{comment}

\section{Empirical Tests}

\subsection{Implementations}

For testing, I have developed the following implementations:
\begin{itemize}
\item \textbf{B\&W}:    The B\&W serial algorithm described by~\citep{Braun2013} adapted from code provided by Braun.
\item \textbf{B\&W+P}:  The B\&W algorithm with only erosion parallelized.
\item \textbf{B\&W+PI}: The B\&W algorithm parallelized using the additional techniques described here, but still using the stack structure.
%\item \textbf{B\&W+TP}: As \textsection\ref{sec:breadth} discusses, parallelizing B\&W by making a parallel task for each tree/stack leads to load imbalancing. At the opposite extreme, making two new tasks every time a tree bifurcates causes unacceptable overhead. This implementation attempts a compromise by creating new parallel tasks only for the first few bifurcations of a tree.
\item \textbf{RB}:      A serial version of the new algorithm.
\item \textbf{RB+P}:    The new algorithm with only erosion parallelized, for comparison against B\&W+P.
\item \textbf{RB+PI}:   The new algorithm using all the parallel techniques described here.
\item \textbf{RB+PQ}:   The new algorithm using all the parallel techniques described here separated by threads.
\item \textbf{RB+GPU}:   The new algorithm using all the parallel techniques described here implemented for use on a GPU.
\end{itemize}

In lieu of further details here, extensively-commented source code for all implementations is available at \url{https://github.com/r-barnes/Barnes2018-Landscape}. All algorithms were targeted to the native architecture of the test machines and compiled using GCC (except where noted) with both full optimizations and ``fast math" enabled. A makefile is provided with the source code.

Minimal effort has been put into low-level optimizations and OpenACC has been used instead of more expressive, but more difficult to use, accelerator frameworks such as CUDA and OpenCL. This is intentional: the code here is meant to be accessible to any HPC-orientated geoscientist and the wall-times reflect this.

\subsection{Machines}

The SummitDev supercomputer of Oak Ridge National Lab's Leadership Computing Facility was used for timing tests. Each node has two 10-core IBM POWER8 CPUs with each core supporting 8 hardware threads (160 threads total). Each node has 500GB DDR4 memory and is attached to 4 NVIDIA Tesla P100 GPUs. Cache sizes are L1=64K, L2=512K, L3=8,192K.
%\item Comet: A supercomputer managed by XSEDE~\cite{xsede}. Each node has two 12-core Intel Xeon E5-2680v3 CPUs with 128GB DDR4 RAM. Each node is attached to 2 NVIDIA K80 GPUs. Cache sizes are L1=32K, L2=256K, L3=30,720K.
%\item Sala: A circa 2010 Thinkpad which the author saved from the bin and uses for the majority of his work. It sports a 4-core Intel i5-480M CPU, 8GB DDR3, and has no GPU. Cache sizes are L1=32K, L2=256K, L3=3,072K.
%\end{itemize}
%Note that, in addition to having more hardware threads (greater parallel potential), the more powerful machines tend to have larger L1 and L2 caches: this can have a significant positive impact on performance.

\subsection{Test Data}

Several sizes of square rasters of elevation were generated. Each cell of the rasters was initialized to a value drawn uniformly from the range $[0,1]$. Seed values were set so that all implementations at a given size used the same data, allowing for safe intercomparison. 

\subsection{Test Iterations}

All tests were run for 120 timesteps to better extract the effect of input size on wall-times. This is sufficient to reach steady-state for small inputs, but additional iterations would be necessary to achieve convergence on larger inputs.

\subsection{Correctness}
The outputs of all of the implementations have been compared and are identical. This suggests that the implementations are correct (or at least all wrong in the same way). The source code includes a script which performs this comparison automatically.

\section{Results \& Discussion}
\label{sec:resultdisc}

\begin{figure}
\centering
\includegraphics[width=\columnwidth]{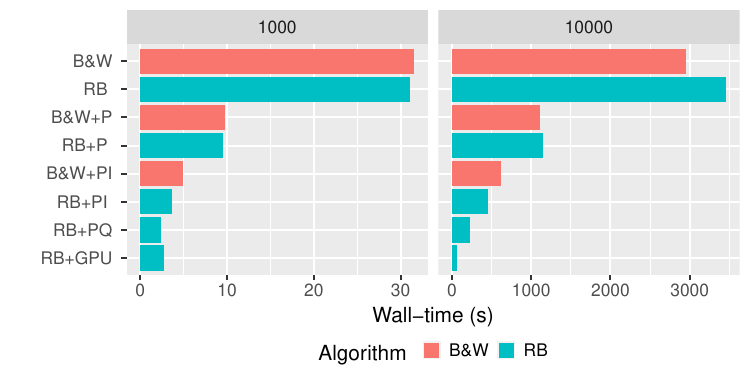} %../src/tests/img_timing_compare.pdf
\caption{Timing comparisons for all implementations for two input sizes. \label{fig:BW-vs-RB-all}}
\end{figure}

\autoref{fig:BW-vs-RB-all} shows the aggregate of the results of the tests below. For the larger dataset, the best parallel implementations of the new algorithm run 13\,x faster than the B\&W serial implementation on a CPU and 43\, faster on a GPU. The performance details of each implementation are discussed below. Note that $x$-axes in the figures are independent.

\subsection{Serial Implementations}

\begin{figure}
\centering
\includegraphics[width=\columnwidth]{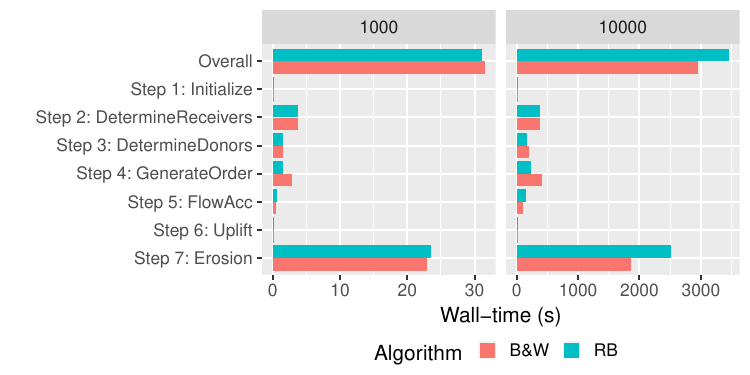} %../src/tests/img_serial_comparison.pdf
\caption{Timing comparisons for the B\&W and RB (serial) implementations. \label{fig:BW-vs-RB}}
\end{figure}

%CW: \printlen[5][in]{\columnwidth}

%NOTES: tied versus united and depth seem to have no effect
%NOTES: BW+P seemed faster than BW+TP

\autoref{fig:BW-vs-RB} compares the wall-times of the B\&W and RB implementations. These serial implementations differ only in whether or not a stack or a queue is used. Note that for the $1000^2$ dataset, the RB implementation is slightly faster, but that it is slower for the $10000^2$ dataset. This indicates that for most models using the breadth-first (queue) traversal should have a negligible impact on speed versus using the depth-first (stack) traversal. As we will see, the breadth-first traversal gives shorter wall-times when parallelism is used. 

\autoref{fig:BW-vs-RB} also shows that the majority of the wall-time (an average of 75\%)
%across both implementations and all dataset sizes
is consumed by the erosion function. Optimizing this is therefore key to improving the efficiency of both algorithms.

\subsection{Parallelizing Erosion}

\begin{figure}
\centering
\includegraphics[width=\columnwidth]{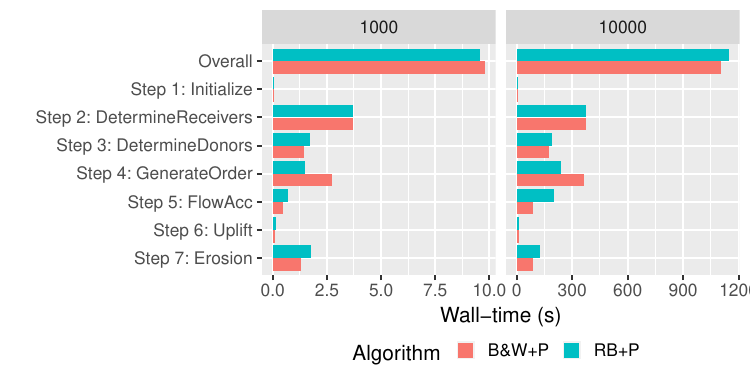} %../src/tests/img_simple_parallel_comparison.pdf
\caption{Timing comparisons for the B\&W+P and RB+P implementations, in which only \textit{Step 7: Erosion} is parallelized. \label{fig:BWp-vs-RBp}}
\end{figure}

The hour-plus wall-time of the larger model in the previous section demonstrates the need for parallelism. The B\&W+P and RB+P implementations address this by parallelizing the erosion function. As shown in \autoref{fig:BWp-vs-RBp}, doing so brings the two algorithms to approximately the same wall-time: 10\,s for the $1000^2$ dataset and 1200\,s for the $10,000^2$ dataset, a 66\% reduction versus serial performance. For these implementations, erosion takes about 25\% of the wall-time.

\subsection{Parallelizing All Steps}

\begin{figure}
\centering
\includegraphics[width=\columnwidth]{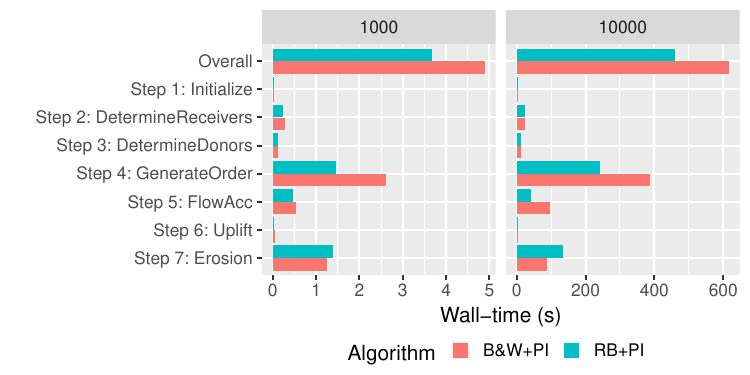} %../src/tests/img_parallel_improved_comparison.pdf
\caption{Timing comparisons for the B\&W+PI and RB+PI implementations, in which all steps of the algorithm are parallelized. \label{fig:BWpi-vs-RBpi}}
\end{figure}

The flat distribution of wall-times across the various steps of the RB+P and B\&W+P implementations shown in the previous section indicate the need to parallelize all of the steps in order to improve efficiency. The results of doing so are shown in \autoref{fig:BWpi-vs-RBpi}. Wall-times are halved versus the previous step and the RB+PI algorithm is now definitively faster.

Profiling of RB+PI shows that synchronization at barriers (discussed earlier) is responsible for most of the time taken by both the erosion and flow accumulation steps.

The construction of the queue/stack (\textit{Step4\_GenerateOrder}) has now emerged as a serious bottleneck. At this point it is still serial in both algorithms. Parallelizing its construction in B\&W by avoiding explicit stack construction via OpenMP tasks did not lead to better performance, though the code for this is provided. Thus, we have reached the limit of parallel gains in the B\&W algorithm. However, it is possible to make further improvements to the RB algorithm.

\subsection{Independent Parallelism}

\begin{figure}
\centering
\includegraphics[width=\columnwidth]{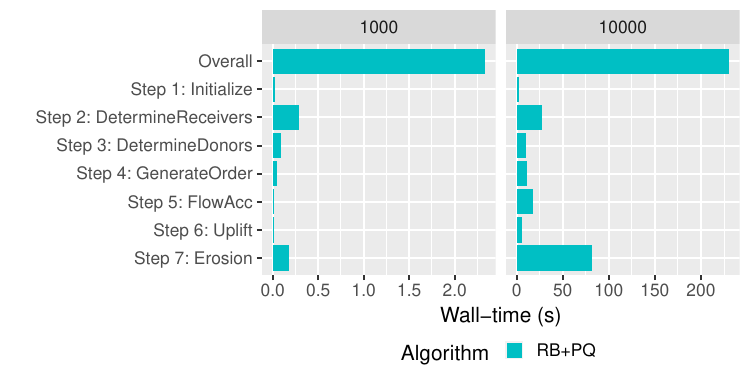} %../src/tests/img_parallel_sep_thread_comet.pdf
\caption{Timings for the RB+PQ implementation, in which separate threads have private memories. \label{fig:RBpq}}
\end{figure}

Queue generation can be parallelized and synchronization barriers eliminated by giving each thread its own private queue in Algorithm~\ref{alg:generate-queue}. Passing this private queue onward to subsequent steps allows several stages of the algorithm to proceed independently. This is implemented in RB+PQ.

As \autoref{fig:RBpq} shows, using private queues halves the run-time again. In this implementation, a barrier has been added after \textit{Step 7: Erosion} which introduces a necessary synchronization delay that accounts for the majority of the wall-time not included within the steps themselves.

\subsection{GPU Implementation}

\begin{figure}
\centering
\includegraphics[width=\columnwidth]{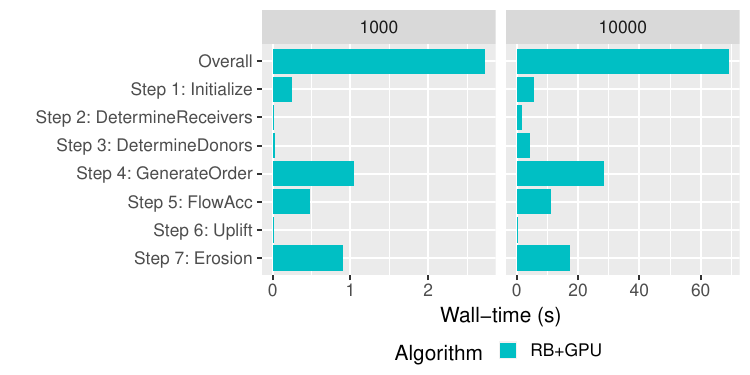} %../src/tests/img_gpu_summitdev.pdf
\caption{Timings for the RB+GPU implementation. \label{fig:RBgpu}}
\end{figure}

\begin{figure}
\centering
\includegraphics[width=\columnwidth]{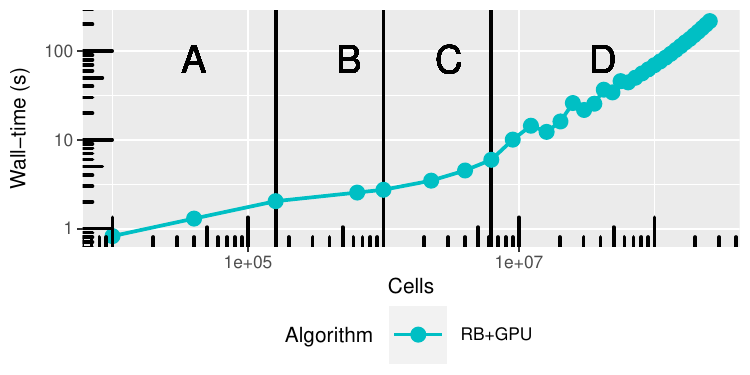} %../src/tests/img_gpu_scaling_summitdev.pdf
\caption{Scaling for the RB+GPU implementation. \label{fig:RBgpuscaling}}
\end{figure}

\begin{table}
\centering
\footnotesize
\begin{tabular}{r|r|r|r}
  &    x &  Edge Length & Cells           \\ \hline
A & 0.33 &          400 & $1.6\cdot10^5$  \\
B & 0.16 &         1000 & $1.0\cdot10^6$  \\
C & 0.42 &         2500 & $6.25\cdot10^6$ \\
D & 0.92 &        16000 & $2.56\cdot10^8$
\end{tabular}
\caption{Scaling of the algorithm for the regions identified in \autoref{fig:RBgpuscaling}. \label{tbl:gpuscaling}}
\end{table}

Further performance can be gained by delegating calculations to a GPU or other many-core processor. Since the GPU's local private memory was too limited for the design of RB+PQ to work, the design of RB+PI was modified by using OpenACC directives and building with the PGI compiler. 

Another possible design was to allocate global memory addressed with a thread-specific index; however, the OpenACC standard does not provide a function for obtaining such an index and the PGI compiler-specific extension \texttt{\_\_pgi\_gangidx()} proved difficult to work with. Similarly, though OpenMP 4.5 provides the \texttt{omp\_get\_team\_num()} function, compiler support at the time of writing was too rudimentary to test this. Therefore, \textit{Step 4: Generate Order} is parallelized by treating the variable \textit{nqueue} in Algorithm~\ref{alg:generate-queue} as an atomic with only a small number of threads used to generate the queue. Future compiler developments will, presumably, enable better design choices.

\autoref{fig:RBgpu} shows the results. For the smaller dataset, the GPU gives no wall-time advantage over the RB+PQ implementation; for the larger dataset, the GPU gives a 3\,x speed-up. The scaling here is notable. A 100\,x data increase in the RB+PQ implementation resulted in a 100\,x increase in wall-time; in contrast, a 100\,x data increase in the RB+GPU implementation resulted in only a 28\,x increase in wall-time. The algorithm is scaling \textit{sublinearly}.

\autoref{fig:RBgpuscaling} and \autoref{tbl:gpuscaling} explore this in-depth. Four distinct behavioural regions can be identified, as shown in the figure. In each region the algorithm's wall-time scales as $O(N^x)$ where $x$ and associated upper bounds of the edge length and number of cells of the region are shown in the table. Performance decreases smoothly throughout region $D$ eventually approaching and passing $x=1$.

The reason for this sublinearity is that the GPU has such a large amount of compute power that the smaller datasets cannot effectively use it all, so only a small fraction is applied to a given problem. The result is wall-times which remain flat as the input size increases.

The GPU has other advantages. Its unused compute power can be used to simultaneously process other models. In a multi-GPU system such as Summitdev, this means many model realizations can be carried out in a short time. GPUs also tend to be more energy-efficient than CPUs, so the net energy, environmental, and monetary costs of doing a given calculation are reduced.

\subsection{Future Improvements}

There are still opportunities to improve GPU performance. \textit{Step 4: Generate Order} is difficult to parallelize because so little computation is done. I have handled this in my implementation by using a small number of threads to atomically increment where indices are stored in the queue. The forthcoming Nvidia Volta's improved atomic performance may help with this situation. Otherwise, compiler improvements may allow a strategy such as RB+PQ to be implemented on GPUs in the near future.

%Neither the RB+PQ and RB+GPU implementations succeed in obtaining the full speed-up suggested by \autoref{fig:avail-parallelism}. This is because although the new design improves the algorithm's floating-point performance, memory is still accessed via indirect addressing which requires the processor to perform inefficient ``gather-scatter" accesses. Future processor designs may help resolve this.

\section{Coda}
The foregoing has detailed algorithmic and methodological approaches to accelerating the modeling of landscape evolution. The result is an algorithm which, in its unoptimized form, matches the performance of the previous state of the art: an algorithm by \citet{Braun2013}. On the CPU, the algorithm runs in less a third the time of the best B\&W parallel implementation. On the GPU, the algorithm runs 43\,x faster than the serial version of the B\&W algorithm, 9\,x faster the best B\&W parallel implementation, and scales sublinearly with input size.

In future work, I will show how the foregoing can be extended to multi-GPU environments in order to quickly solve inverse problems and perform sensitivity analysis on landscape evolution models.

Complete source code and tests are available at \url{https://github.com/r-barnes/Barnes2018-Landscape}.

\section{Acknowledgments}
This work was supported by the Department of Energy's Computational Science Graduate Fellowship (Grant No.\ DE-FG02-97ER25308), the National Science Foundation's Graduate Research Fellowship, and an SC travel grant.

Empirical tests and results were performed on Summitdev, which is a prototype machine for the forthcoming Summit supercomputer managed by Oak Ridge National Laboratory's Leadership Computing Facility, and XSEDE's Comet supercomputer~\citep{xsede}, which is supported by the National Science Foundation (Grant No.\ ACI-1053575). 

Some of the techniques used in the paper were learned at the CSGF Program Review's ``Mini-GPU Hackathon" led by Fernanda Foertter, Thomas Papatheodore, Adam Simpson, Verónica Vergara Larrea, Mark Berrill, and Matthew Norman. Jack DeSlippe and Thorsten Kurth helped with an unused OpenMP implementation at an LBNL KNL Hackathon. Mat Colgrave of the PGI Compiler Group found bugs in both my code and the PGI compiler. Kelly Kochanski provided helpful discussion and ideas.

In-kind support was provided by
Lorraine B., %arnes
Myron B., %arnes
Hannah J., %ohlas
Kelly K., %ochanski
Lydia M., %cAnerny
Myron M., %aki
John O., %rrison
and Jerry W.

\section{Bibliography}

{\footnotesize
  \bibliographystyle{elsarticle-harv} 
  \bibliography{refs}   % expects file "myrefs.bib"
}

\end{document}